\documentclass{kapproc} 

\RequirePackage{graphicx}%
\RequirePackage{epsf}%
\input{psfig}

\upperandlowercase
\let\footnote\savefootnote
\let\footnotetext\savefootnotetext 
\let\footnoterule\savefootnoterule 
\kluwerbib 

\begin{document}

\articletitle{Halo Mass Function}
\articlesubtitle{Low-mass Stars in Deep Fields}

\chaptitlerunninghead{Halo Mass Function}

 \author{Wolfgang Brandner}
 \affil{Max-Planck-Institut f\"ur Astronomie, K\"onigstuhl 17, D-69117 Heidelberg, Germany}
 \email{brandner@mpia.de}

 \begin{abstract}
Deep fields are unique probes of the Galactic halo. With a limiting 
magnitude of, e.g., I = 30 mag, all stars down to the hydrogen-burning 
limit are detected out to distances of 10 kpc, while stars with
0.5 solar masses could be traced out to distances of 400 kpc. Thus deep
fields provide an opportunity to study both the structure of the Galactic
halo and the mass function of population (Pop) II and Pop III stars. I 
summarize previous work on the Galactic stellar halo utilizing the Hubble 
Deep Fields North and South, supplemented by a preliminary analysis of 
the recently released Hubble Ultra-Deep Field.
 \end{abstract}

\section{Motivation}

The structure and stellar content of the Galactic Halo directly points
back to the formation of the Milky Way. Initial studies naturally focused
on the Pop II main-sequence stars in the solar neighbourhood. 
These limits have been gradually extended towards fainter (i.e. lower mass 
and/or more distant) halo stars, including evolved stars, white dwarfs and
a search for the baryonic dark matter component (e.g.\ Paczynski 
\cite{paczynski86}, Alfonso et al.\ \cite{alfonso03}, and references therein).

Studies of stellar populations in deep fields aim at addressing the following
issues:

\begin{itemize}
\item
Stellar Content and Origin of the Galactic Halo
\item
(Global) Structure of the Galactic Halo
\item
Search for a (still self-luminous) baryonic component of the ''dark matter''
halo
\item
Fraction of Pop {\sc III} stars in the Galactic Halo, in particular
stars with [Fe/H] $< -4$ (see, e.g., Oey 2003)
\end{itemize}

\subsection{Galactic Structure}

The stellar population of the Milky Way is arranged into three
dynamically distinct components.  Bahcall \& Soneira (1981, 1984) 
provided evidence for a thin disk of Pop I stars, and a halo
composed of Pop II stars. Gilmore \& Reid (1983) and Gilmore (1984) 
identified the
thick disk as a third distinct component. Figure 1 illustrates the 
3-component composition of the Milky Way stellar population as
evidenced by SDSS data (Chen et al.\ 2001). Chen et al.\ (2001),
Larsen \& Humphreys (2003), and Lemon et al.\ (2004) all derive a flattened 
halo density distribution
with c/a $\approx$ 0.55, following a density law $\rho {\rm (r)} \propto
{\rm r}^{-2.5 \pm 0.3}$.

\begin{figure}[ht]
\centerline{
\psfig{file=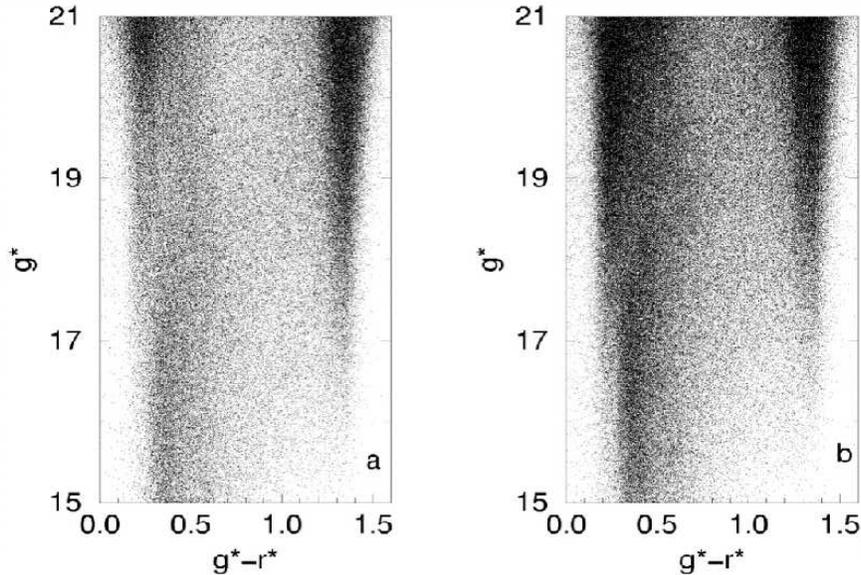,width=11.5cm,angle=270}
}
\caption{Colour-Magnitude Diagram based on SDSS data for high Galactic latitudes 
of b$ > +48^\circ$ (left) and b$ < -48^\circ$ (right). The three component
structure of the Milky Way with the halo at $g^*-r^* \approx 0.2$\,mag, the 
thick disk at $g^*-r^* \approx 0.33$\,mag and a turn-off magnitude of 
$g^* \le 18$\,mag, and the thin disk at  $g^*-r^* \approx 1.4$\,mag is clearly 
evident (Chen et al.\ 2001).}
\end{figure}

It should be noted, however, that the results quoted above are based on the
assumption of a uniform, global Salpeter-type (Salpeter 1955) luminosity 
function.

\subsection{Subdwarf Luminosity Function}

Over the past 15 years, a number of studies with increasingly fainter magnitude
limits and larger field coverage aimed at the derivation of the
subdwarf luminosity function. Gould et al.\ (1998) report a discrepancy
between the luminosity functions for the inner and the outer halo. If
confirmed, this would constitute a case of a radially varying mass function
in the halo of the Milky Way. A more recent study by Digby et al.\ (2003),
however, reaching out to M$_{\rm V} \le $12\,mag for heliocentric distances
of 2.5\,kpc, did not find any indication for a variation
in the mass function (see Figure 2).

\begin{figure}[ht]
\centerline{
\psfig{file=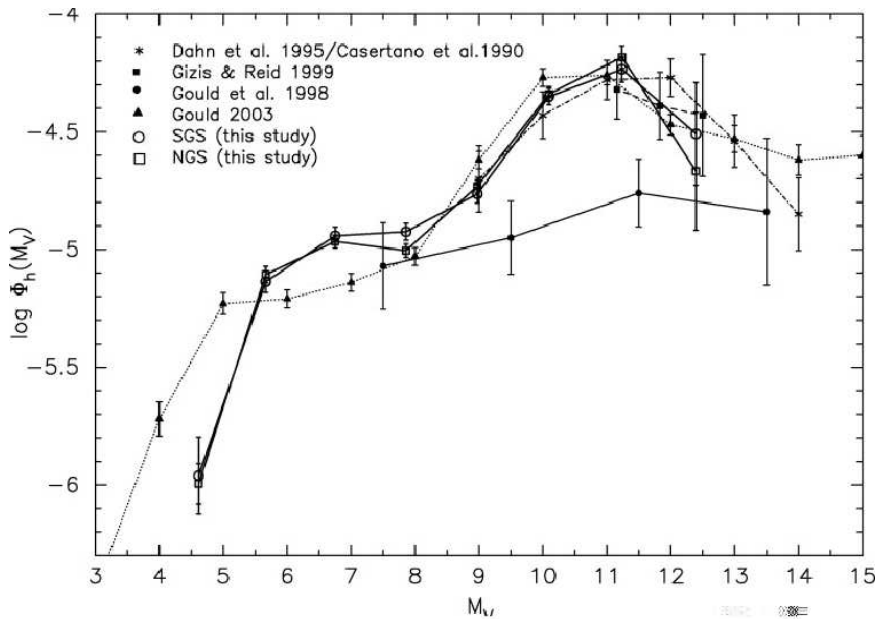,width=11.5cm,angle=270}
}
\caption{Compilation of estimates of the luminosity function for subdwarfs from Digby et al.\
(2003).}
\end{figure}

\section{Hubble Deep Fields and the Halo IMF}

The Hubble Deep Fields North (HDF-N) and South (HDF-S) were observed in December
1995 and October 1998,
respectively, with the aim to obtain the most detailed view of distant
field galaxies in order to study various aspects of galaxy evolution and
cosmology based on galaxy number counts, luminosity functions and morphology 
(Williams et al.\ 1996, 2000). In order to minimize contamination by
Galactic stars, the field were centered far from the Galactic plane at
latitudes of $b = +54^\circ$ and $b = -49^\circ$, respectively.
The Hubble Ultra-Deep Field (UDF), obtained in
late 2003, represents an even deeper image of a high-Galactic latitude
field (Beckwith et al.\ 2004). Table 1 summarizes the basic properties
of HDF-N/S and UDF.

\begin{table}[ht]
\caption{\label{tab1}Properties of HDF-N/S$^a$ and UDF$^b$}
\begin{tabular*}{\textwidth}{@{\extracolsep{\fill}}clcccc}
\sphline
\it Field&\it l &\it b & \it Filters & \it 5$\sigma$ limit & \it \# of \cr
\it      &\it   &\it   & \it         & \it (AB mag)        & \it MS stars \cr
\sphline
HDF-N & 126$^\circ$ & +54$^\circ$ & UBVI & I $\approx$ 29.1 & 9 \cr
HDF-S & 328$^\circ$ & -49$^\circ$ & UBVI & I $\approx$ 29.1 & 29 \cr
UDF   & 223$^\circ$ & -54$^\circ$ & BVIz & I $\approx$ 30.5 & 9 (11) \cr
\sphline
\end{tabular*}
\begin{tablenotes}
$^a$ Williams et al.\ 1996, 2000

$^b$ Beckwith et al.\ 2004
\end{tablenotes}
\end{table}

Figure 3 shows colour-magnitude diagrams for the UDF based on the
ACS/WFC source catalogue version 1. 

\begin{figure}[ht]
\centerline{
\psfig{file=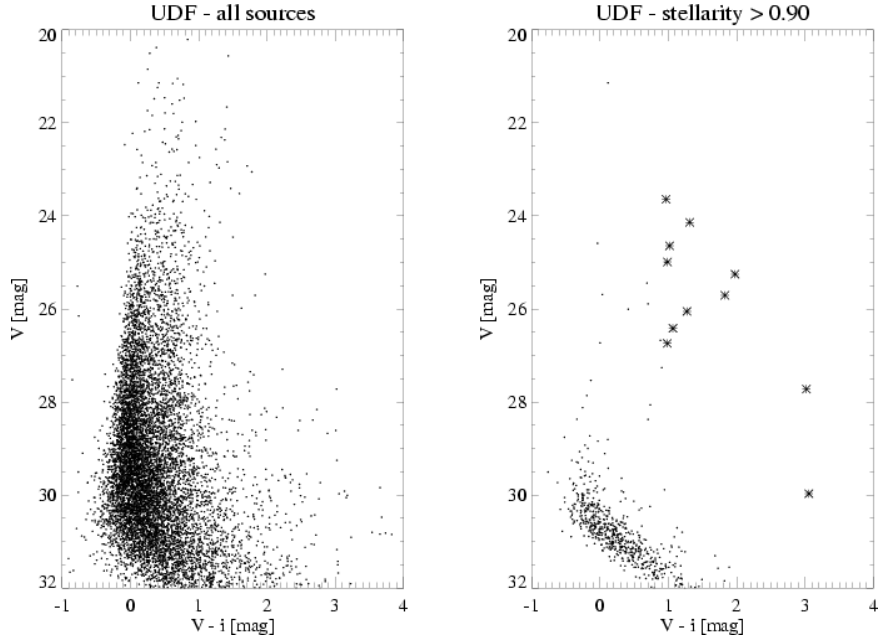,width=11.5cm,angle=270}
}
\caption{Colour-magnitude diagram for UDF. The diagram on the
left shows all sources, whereas the diagram on the right only
includes sources with a stellarity index $>$ 0.90. Candidate
main-sequence stars are marked by ``star'' symbols.}
\end{figure}

A pioneering study of the {\it stellar} content of HDF-N/S was carried out
by the late Rebecca Elson and collaborators (Elson et al.\ 1996; Johnson
et al.\ 1999). The number of main-sequence stars in both fields (see
Figure 4) turned out to be in good agreement with models. In particular,
they found no evidence for an extended, cD-like stellar halo around the Galaxy.
Johnson et al.\ (1999) argue that the compact blue sources identified
in HDF-S most likely constitute unresolved (extragalactic) star forming
regions, rathern than Pop III white dwarfs.

\begin{figure}[ht]
\centerline{
\hbox{
\psfig{file=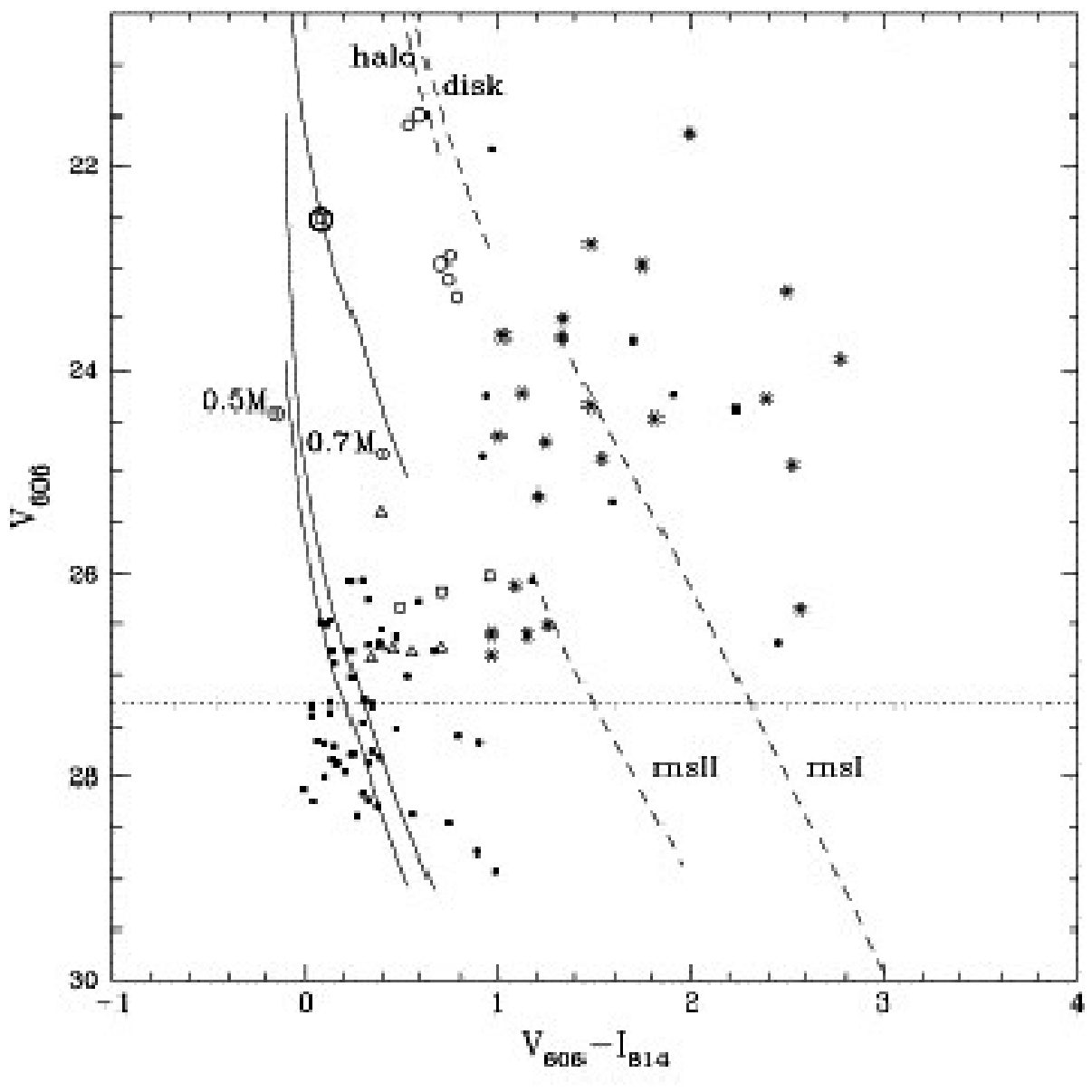,width=6.0cm,angle=270}
\psfig{file=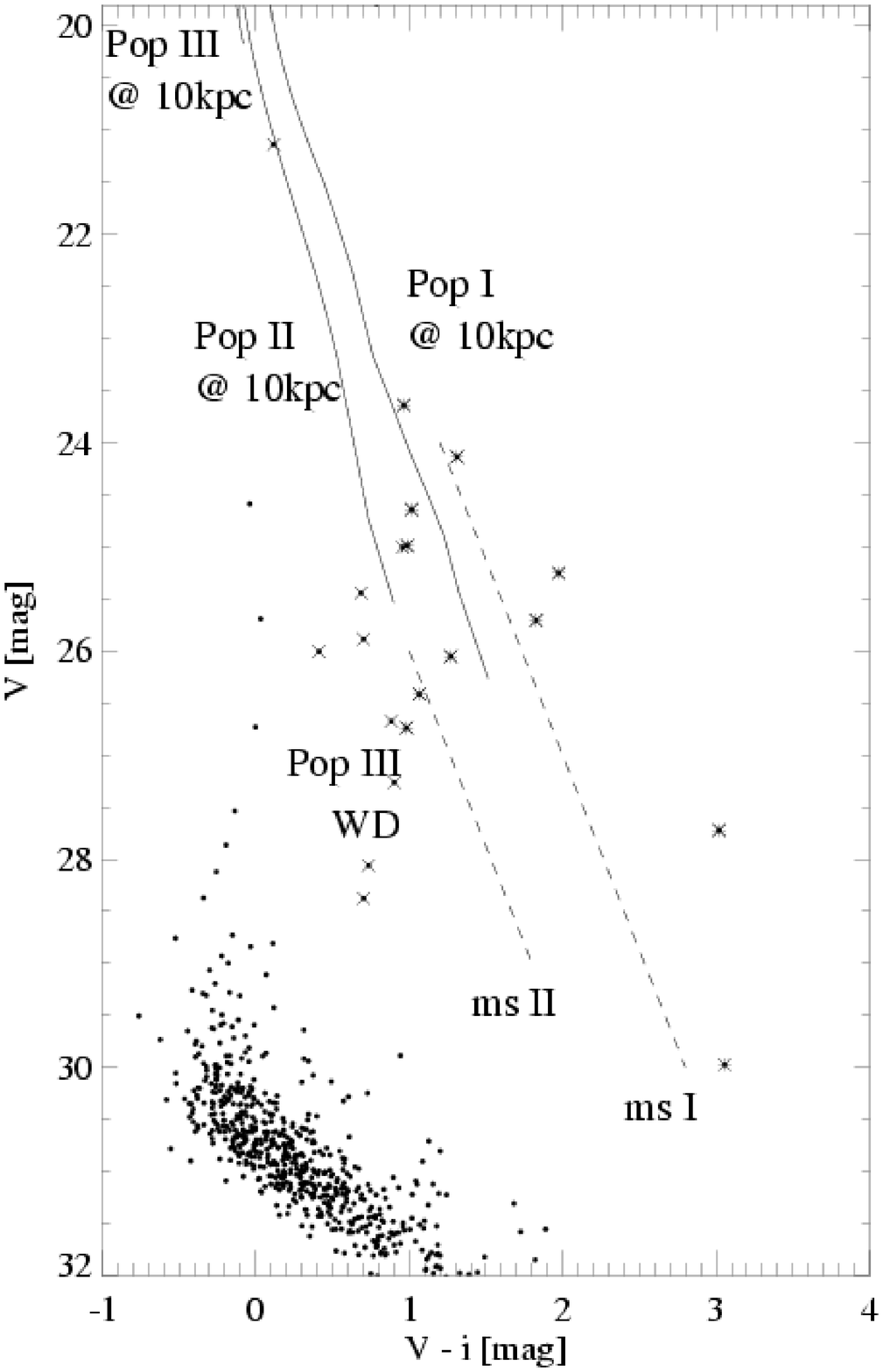,width=6.0cm,height=7.6cm,angle=180}
}
}
\caption{Colour-magnitude diagram for HDF N and S (left) and UDF (right).
The locations of Pop I, II and III main-sequence stars (for a distance
of 10\,kpc), and cooling curves for white dwarfs with masses of 0.5 and
0.7\,M$_\odot$ are shown as well.}
\end{figure}

In a related study, Ibata et al.\ (1999) and Kalirai et al.\ (2004)
analysed deep, multi-epoch HST images of the inner halo in the direction
of M4 (l = 351$^\circ$, b = 16$^\circ$). Down to a limiting magnitude of 
V = 29\,mag, the observed numbers of thin-disk, thick-disk and Pop II 
halo white dwarfs are in good agreement with Galactic models. Based on
a Chabrier IMF (Chabrier 1999) they expected to find 2 to 3 Pop 
III white dwarfs, but found none. While there is no need to invoke 
Pop III white dwarfs in order to fit the data, the small number
statistics also does not yet rule out the existence of Pop III
white dwarfs.

\section{Open Questions}

The following questions still need to be answered:

\begin{itemize}
\item
Are all the red, low-mass candidates identified in HDF-N/S and UDF indeed stars?
\item
What is the nature of the compact blue sources? Are they Pop III white dwarfs or unresolved star forming regions?
\item
What is the exact shape of the outer halo luminosity function?
\end{itemize}

To address these questions, a spectroscopic follow-up of the 
brighter Pop II and Pop III candidates should be carried out. Spectroscopy
of the fainter candidates is a science case for the next generation of 
Extremely Large Telescopes with apertures in the range 30\,m to 100\,m.

2nd epoch imaging of HDF-N (S) in Cycle 14 (15) would provide an epoch
difference of 10 (8) years. Assuming an astrometric precision of 5\,mas,
it would be possible to confirm proper motions as small as 0.5\,mas/yr. This
should be sufficient to distinguish Pop II and Pop III white dwarfs from
unresolved, extragalactic star forming regions, and might lead to the
first identification of Pop III objects.

\begin{acknowledgments}
Discussions with Dimitrios Gouliermis and Bertrand Goldman are gratefully
acknowledged.

\end{acknowledgments}

\end{document}